\documentstyle[12pt]{article}
\def\lsim{\mathrel{\rlap {\raise.5ex\hbox{$ < $}}
{\lower.5ex\hbox{$\sim$}}}}

\newcommand{\pr}{\paragraph{}}
\newcommand{\be}{\begin{equation}}
\newcommand{\ee}{\end{equation}}
\newcommand{\bea}{\begin{eqnarray}}
\newcommand{\nn}{\nonumber}
\newcommand{\eea}{\end{eqnarray}}
\newcommand{\nd}[1]{/\hspace{-0.6em} #1}
\newcommand{\nk}{\noindent}
\def\gappeq{\mathrel{\rlap {\raise.5ex\hbox{$>$}}
{\lower.5ex\hbox{$\sim$}}}}

\def\lappeq{\mathrel{\rlap{\raise.5ex\hbox{$<$}}
{\lower.5ex\hbox{$\sim$}}}}

\begin{document}

\begin{titlepage}
\begin{flushright}
CERN-TH/95-336 \\
CTP-TAMU-04/95 \\
ACT-02/95 \\
OUTP-96--05P \\
\end{flushright}
\begin{centering}
\vspace{.1in}
{\large {\bf Quantum
Decoherence in a Four-Dimensional Black Hole Background }
} \\
\vspace{.2in}
{\bf John Ellis$^{a} $},
{\bf N.E. Mavromatos$^{b,\diamond}$},
{\bf D.V. Nanopoulos}$^{c}$ \\
and \\
{\bf Elizabeth Winstanley$^{b}$}
\\
\vspace{.03in}
\vspace{.1in}
{\bf Abstract} \\
\vspace{.05in}
\end{centering}
{\small  We display a logarithmic divergence in the density
matrix of a scalar field in the presence of an Einstein-Yang-Mills
black hole in four dimensions. This divergence is related to a
previously-found logarithmic divergence in the entropy of the
scalar field, which cannot be absorbed into a renormalization
of the Hawking-Bekenstein entropy of the black hole. As the
latter decays, the logarithmic divergence induces a non-commutator
term $\nd{\delta H}\rho$ in the quantum Liouville equation for the
density matrix $\rho$ of the scalar field,
leading to quantum decoherence. The order of magnitude of
$\nd{\delta H}$ is $\mu^2/M_P$, where $\mu$ is the mass of
the scalar particle.}
\vspace{0.2in}
\nk $^a$ Theory Division, CERN, CH-1211, Geneva, Switzerland,  \\
$^b$
University of Oxford, Dept. of Physics
(Theoretical Physics),
1 Keble Road, Oxford OX1 3NP, United Kingdom,   \\
$^{c}$ Center for
Theoretical Physics, Dept. of Physics,
Texas A \& M University, College Station, TX 77843-4242, USA
and Astroparticle Physics Group, Houston
Advanced Research Center (HARC), The Mitchell Campus,
Woodlands, TX 77381, USA. \\
$^{\diamond}$ P.P.A.R.C. Advanced Fellow.

\vspace{0.01in}
\begin{flushleft}
CERN-TH/95-336 \\
CTP-TAMU-04/95 \\
ACT-02/95 \\
OUTP-96--05P \\
January  1996\\
\end{flushleft}
\end{titlepage}
\newpage

\section{Introduction and Summary}
\pr
There is currently much debate whether microscopic black holes
induce quantum decoherence at a microscopic level. In particular,
it has been suggested~\cite{Hawking}
that Planck-scale black holes and other
topological fluctuations in the space-time background cause
a breakdown of the
conventional $S$-matrix description of asymptotic particle
scattering in local quantum field theory, which should be
replaced by a non-factorizable superscattering operator
$\nd{S}$ relating initial- and final-state density matrices:
\be
\rho_{out} = \nd{S} \rho_{in}
\label{hawk}
\ee
It has further been pointed out that, if this suggestion
is correct, there must be a modification of the usual
quantum-mechanical time evolution of the wave
function, taking the form of a modified Liouville equation
for the density matrix~\cite{EHNS}:
\be
\partial _t \rho =  i [\rho, H] + \nd{\delta H}\rho
\label{ehns}
\ee
The extra term in (\ref{ehns}) is of the form generally
encountered in the description of an open quantum-mechanical
system, in which observable degrees of freedom are coupled
to unobservable components which are effectively
integrated over, which may evolve from a pure state to a
mixed state with a corresponding increase in entropy. The
necessity of a mixed-state description is generally accepted
in the presence of a macroscopic black hole, but is far from
being universally accepted in the case of microscopic virtual
black-hole fluctuations.

\pr
Three of us have been analyzing
the possibility of quantum decoherence in a non-critical
formulation of string theory \cite{EMN}, in which the two-dimensional
string black hole solution \cite{Wbh}
is used as an example of a quantum
fluctuation in space time. We exhibited in this model extra
logarithmic singularities, associated with transitions between
different conformal field theories on the world sheet, which
induced an extra term in the quantum Liouville equation of the
form conjectured in (\ref{ehns}). This derivation of
(\ref{ehns}) was, however, incomplete, in that two dimensions
are not the same as four, and even in two dimensions one does not
have a complete non-perturbative formulation of string theory
that includes transitions between different conformal field
theories\footnote{Progress in this direction is being made
via studies of duality and conifolds \cite{dualcon}, with
considerable similarity to the approach taken in \cite{EMN}: see
\cite{EMNdual}.}
In parallel, two of us have been
studying a scalar field in the presence of a four-dimensional
Einstein-Yang-Mills black-hole background \cite{MW},
and have demonstrated that its entropy exhibits a
logarithmic divergence that cannot be absorbed into a simple
renormalization of the Bekenstein entropy of the black hole.
This can be understood as a quantum reflection of the
entanglement of the external scalar field with internal
black hole properties that alter with its radius.

\pr
The purpose of this paper is to show that this logarithmic
divergence, which also appears in the partition function $Z$ but
not at the same order in the Hamiltonian, induces
an extra non-commutator term into the quantum Liouville
equation for the density matrix of the scalar field. This
follows from the identification of the renormailization scale
with time, as in the two-dimensional non-critical string~\cite{EMN},
and
constitutes the first example of a derivation of equation
(\ref{ehns}) in four dimensions. We argue that, once one
allows for the quantum decay of the black hole, the
logarithmic divergence found in the previous static case
becomes a new source of time dependence for the scalar
field that leads to a monotonic
 increase in entropy, and hence cannot be
absorbed within the conventional Liouville equation.
The order of magnitude that we find for the extra term is
\be
\nd{\delta H}~~\simeq~~\frac{\mu^2}{M_P},
\label{magnitude}
\ee
where $\mu$ is the mass of a quantum of the scalar field.
The estimate (\ref{magnitude}) is the maximum that had been
indicated by previous string black hole studies
\cite{EMN}, and is not very far from the sensitivity of
present experimental searches \cite{CPLEAR}.

\section{Entropy, Hair and Logarithmic Divergences}
\pr
As a first step in this analysis, we now review
relevant previous work and recall some of the ingredients
we use. The observation that quantum black holes must be
described by mixed quantum states goes back to the work of
Bekenstein~\cite{HB} and Hawking~\cite{decay}, who
showed at the tree level that a generic black
hole has non-zero entropy related to the area of its event
horizon:
\be
S = \frac{1}{4G_N}A
\label{entr}
\ee
This entropy represents the number of
quantum black hole states that are not distinguishable by
traditional measurements using the varieties of hair known
in conventional local four-dimensional quantum field theories.
It is known that string theories possess an infinite number
of local symmetries, and we have suggested that a corresponding
infinite set of conserved quantum numbers, baptized $W$-hair
\cite{measW},
characterize string black hole states, at least in two dimensions.
We have also pointed out \cite{states}
that the number of string black hole
states grows in the same way as the logarithm of the
Hawking-Bekenstein entropy (\ref{entr}),
and suggested ways in which the
$W$ quantum numbers could in principle be measured and used to
distinguish these string black hole states \cite{measW}.

\pr
However, we have also pointed out that a complete set of these
measurements is not possible in practice, and have gone on to
suggest that the incompletenes  of the $W$-hair measurements
provides a measure of the loss of information associated with
a string black hole \cite{EMN}. The observable low-energy degrees of
freedom are entangled by $W$ symmetry with unobservable
internal states of the black hole. Integration over these
unobserved degrees of freedom provides a truncated subtheory
which possesses logarithmic divergences absent in a conformal
field theory. The latter would represent string theory in a
fixed classical background, in which scattering is described
by a conventional $S$ matrix. The extra logarithmic
divergences are associated with fluctuations in the black-hole
background, which require transitions between different
conformal field theories. Two-dimensional examples of of these
divergences
are provided by string world-sheet monopoles and instantons,
which represent the creation and mass shifts of black holes,
respectively \cite{EMN}. After identification of the renormalization
scale appearing in the logarithms with a time-like Liouville
field, and thence with the target time variable, we have been
able to derive (\ref{ehns}) with an explicit representation
for the non-commutator term.

\pr
We now point out the analogies with the study \cite{MW} of
a scalar field coupled to an Einstein-Yang-Mills (EYM) black hole
in four dimensions. This study was purely in the context of
quantum gravity, with no string degrees of freedom. On the
other hand, the presence of the gauge field endows the EYM
black hole with additional gauge hair (c.f., the $W$ hair of
the two-dimensional string black hole) that is entangled with
the scalar field via gauge interactions (c.f., non-diagonal
$W$ generators). The renormalizability of the EYM field theory
enables the partition function and entropy of the
four-dimensional black hole to be calculated
with the inclusion of quantum corrections, with the results~\cite{MW}:
\bea
&~& -lnZ  \equiv F = -\frac{2\pi^3}{45{\hat \epsilon}}
\frac{r_h^4 (1-2{\hat m}'_h)^{-2}}{\beta ^4} + \nn \\
&~&\{ \frac{4}{45}r_h^3 \frac{\pi^3}{\beta ^4}
\frac{2-4{\hat m}'_h + {\hat m}''_h }{(1-2{\hat m}'_h )^3}
-\frac{1}{6}r_h^3\frac{\pi ^2}{\beta ^2} \mu ^2 \frac{1}{1-
2{\hat m}'_h} \}log~{\hat \epsilon}
\, : \qquad
\beta \equiv  \frac{4\pi r_h e^{\delta _h} }{1-2{\hat m}'_h}
\label{partfunct}
\eea
and
\bea
&~& S  =  S_{BH} + S_q =
\frac {1}{4G_{N,0}}A +  \nn \\
&~&[\frac{(1-2m_{h}')e^{-3\delta _{h}}}{360 \pi r_{h}\epsilon }
\frac {1}{4} A
-\left\{ \frac{1}{180}
\left( 2 -4m_h ' + \frac{m_h''}{r_h} \right)
e^{-3\delta _h} - \frac{1}{12} r_h^2 \mu ^2 e^{-\delta _h}  \right\}
\log \left( \frac {\epsilon }{r_{h}} \right) ]
\nn \\
 &~& \equiv  \frac{1}{G_{N,ren}}\pi r_{q}^{2}
\label{NES}
\eea
where
$S_{BH}$ is the tree-level Bekenstein-Hawking entropy
(\ref{entr}), and the $S_q$ are the quantum corrections,
indicated by square brackets.
The subscript $h$ denotes quantities at the
horizon of the black hole, $r_h$
is the horizon radius, the symbols $\hat X$ denote ratios
$X/r_h$, primes denote differentiation
with respect to the radial coordinate $r$,  $m$ is the
mass function, $\delta $ is defined in ref. \cite{MW} and
will be given explicitly below, and $\epsilon$ is
a small, positive fixed distance which will play the r\^ole of
an ultra-violet cut-off. Its presence is associated with
the `brick wall' boundary condition~\cite{thooft} for the
wavefunction of the (scalar) matter fields in the
black hole background.

\pr
The first term in the
quantum corrections to the entropy (\ref{NES})
can be absorbed into the bare lowest-order Hawking-Bekenstein
entropy (\ref{entr}) via a renormalization of Newton's
constant $G_{N}:~G_{N_0} \rightarrow G_{N,ren}$. However,
the second quantum correction cannot simply
be absorbed into a bare parameter in this way,
and therefore corresponds to a new
effect beyond the reach of the conventional renormalization
programme, as we discuss in the next section.
Formally, it may be absorbed
into a ``quantum'' horizon area, corresponding to
a ``quantum'' radius $r_{q}$, as we shall discuss in section 4.

\section{Hamiltonian of the Scalar Field
in an EYM Black-Hole Background}
\pr
In order to gather information needed for the interpretation
of the logarithmic divergence in the
partition function (\ref{partfunct}) and in the
entropy (\ref{NES}) \cite{MW},
we now compute the Hamiltonian of a
scalar field of mass $\mu$ described by the Lagrangian
\be
{\cal {L}} =-\frac {1}{2} {\sqrt {-g}} (
g^{\mu \lambda }\partial _{\mu } \phi \partial _{\lambda } \phi
+\mu ^{2} \phi ^{2} )
\label{Lag}
\ee
which satisfies the Klein-Gordon equation in an EYM
black-hole background described by the metric
\be
ds^{2} = -e^{\Gamma } dt^{2} + e^{\Lambda } dr^{2}
+r^{2} (d\theta ^{2} +\sin \theta d\varphi ^{2})
\label{metric}
\ee
where the forms of the metric functions $\Gamma, \Lambda$ are
discussed in \cite{MW} and below.
It is easy to verify that the canonical momentum $\pi$ which is
conjugate to $\phi$ is given in this case by
\be
\pi =\frac {\partial {\cal {L}}}{\partial (\partial _{t} \phi )}
=-{\sqrt {-g}} g^{t\nu }\partial _{\nu } \phi
=e^{\frac{1}{2}(-\Gamma +\Lambda )} r^{2} \sin \theta \partial _{t}\phi
\label{pi}
\ee
Defining the Hamiltonian density $\cal H = \pi \partial_t \phi -
\cal L$ in the normal way, we arrive at the following
expression for the Hamiltonian $H$:
\bea
H &=& \int dr d\theta d\varphi \cal H\,:\,
\nn \\
\cal {H} & = &
\frac {1}{2} e^{1/2(\Gamma +\Lambda )} r^{2} \sin \theta
\left( e^{-\Gamma } (\partial _{t} \phi )^{2}
+e^{-\Lambda }(\partial _{r}\phi )^{2} + \right. \nn \\ & &
\left.
\frac {1}{r^{2}} (\partial _{\theta }\phi )^{2}
+\frac {1}{r^{2}} {\mbox {cosec}} ^{2} \theta (\partial _{\varphi } \phi )^{2}
+\mu ^{2} \phi ^{2} \right)
\label{Ham}
\eea
which we now evaluate.

\pr
To evaluate (\ref{Ham}), we first expand $\phi$ in normal modes
\be
\phi (t,r,\theta ,\varphi ) =e^{-iEt} f_{El}(r) Z_{El}(\theta, \varphi )
\label{modes}
\ee
where $Z_{lm}$ is a real spherical harmonic \cite{sred}, and
$f_{El}$ satisfies the radial equation
\be
e^{\Lambda }E^{2} f_{El} +\frac {1}{r^{2}}
e^{-\frac {1}{2}(\Gamma +\Lambda )} \frac {d}{dr}
\left[ e^{\frac {1}{2} (\Gamma -\Lambda )}r^{2}
\frac {d}{dr} f_{El} \right]
-\left( \frac {l(l+1)}{r^{2}} +\mu ^{2} \right) f_{El} =0
\label{radial}
\ee
which we solve with the `brick-wall' boundary conditions~\cite{thooft}
$\phi=0$ at $r=r_h+\epsilon$ and $r=L:L>>r_h$.
Here $r_h$
is the horizon radius of the black hole, $\epsilon$ is
a small, positive fixed distance which will play the r\^ole of
an ultra-violet cut-off, and $L$ is an infra-red cut-off.
Introducing annihilation and creation operators $a$ and
$a^{\dagger}$ for each normal mode (\ref{modes}) in the
usual way, we arrive at the following expression for the
Hamiltonian:
\be
H=- \frac {1}{2} \sum _{E,l,m} \left(
a_{El}a_{El}^{\dagger }+a_{El}^{\dagger }a_{El}
\right) \left[
\frac {1}{2} E + E^{2} J_{El} \right] .
\label{Hami}
\ee
where all the dependence on $\epsilon $ is contained in the last
factor in (\ref{Hami}) and
\be
J_{El} =
\int _{r_{h}+\epsilon }^{L} r^{2}
e^{\frac {1}{2}(\Gamma +3\Lambda )} f_{El}^{2} dr.
\label{jel}
\ee
Before we estimate this integral, a few comments are in order
concerning (\ref{Hami}).

\pr
First, the form of the Hamiltonian
ensures that the Fock basis states are eigenstates of energy,
so that the Hamiltonian can be written in the alternative form
\bea
H &= &\sum _{\mbox {states}}
\langle {}^{1}n_{k_{1}}, {}^{2} n_{k_{2}}, \ldots
{}^{j}n_{k_{j}} |H| {}^{1}n_{k_{1}}, {}^{2} n_{k_{2}}, \ldots
{}^{j}n_{k_{j}} \rangle  \nn \\ & & \times
| {}^{1}n_{k_{1}}, {}^{2} n_{k_{2}}, \ldots
{}^{j}n_{k_{j}} \rangle \langle  {}^{1}n_{k_{1}}, {}^{2} n_{k_{2}},
\ldots {}^{j}n_{k_{j}} |
\eea
where the expectation values are, from (\ref{Hami}),
\bea
\langle {}^{1}n_{k_{1}}, {}^{2} n_{k_{2}}, \ldots
{}^{j}n_{k_{j}} |H| {}^{1}n_{k_{1}}, {}^{2} n_{k_{2}}, \ldots
{}^{j}n_{k_{j}} \rangle & = &
-\frac {1}{2}  \sum _{\mbox {modes}}
 \left[
\frac {1}{2} E + E^{2} J_{El} \right]
\nn \\ & &
-\frac {1}{2} \sum _{i} {}^{i}n  \left[
\frac {1}{2} E_{i} + E_{i}^{2} J_{E_{i}l_{i}} \right] .
\label{expect}
\eea
The first term in (\ref{expect}) is state-independent and contributes
an infinite constant to the partition function
$Z={\mbox {Tr}}e^{-\beta H}$. We shall ignore this
factor from here on,
since it does not contribute to correlation functions.
The second term in (\ref{expect}) will be compared
later in this section to that used in
\cite{MW} to calculate the partition function.

\pr
In order to estimate the integral $J_{El}$, we shall use
\cite{MW} the WKB approximation for the $f_{El}$:
\bea
f_{El} (r)&=&\frac {1}{r} e^{-\frac {1}{2}(\Gamma -\Lambda )}
\frac {A_{El}}{{\sqrt {K_{El}}}(r)}
\sin \left( \int _{r_{h}+\epsilon }^{r}
K_{El}(r') dr' \right)
\label{wkb}
\eea
where the radial wave number $K$ is given by
\be
K_{El}(r)= e^{\Lambda } \left[ E^{2} -\frac {l(l+1)}{r^{2}}
e^{-\Lambda } -\mu ^{2} e^{-\Lambda } \right] ^{\frac {1}{2}}
\label{waveno}
\ee
The constant $A_{El}$ in each case is determined from the
normalization conditions for the $f$'s, namely
\be
\int _{r_{h} +\epsilon }^{L} r^{2}
e^{-\frac {1}{2}(\Gamma +\Lambda )} f_{El}^{2} dr
= \frac {1}{2E}
\label{normal}
\ee
and the boundary condition
\be
f_{El}(L)=0
\ee
implies that
\be
\int _{r_{h}+\epsilon }^{L} K_{El}(r') dr' =n\pi
\qquad
{\mbox {for some integer $n$}}.
\ee
Next introduce a dimensionless co-ordinate $x$ given by
\be
x=\frac {r}{r_{h}}
\label{xdef}
\ee
and the rescaled function ${\hat {K}}$ as follows:
\be
{\hat {K}}_{El}=r_{h}K_{El}
\label{khat}
\ee
and
\be
 {\hat \epsilon } = \frac{\epsilon}{r_h}.
\label{ehat}
\ee
We decompose the integral in (\ref{normal}) as
$A_{El}^{2}(I_{1}+I_{2})$, where
$I_{1}$ is the contribution to the integral for $x$ very close
to one, and $I_{2}$ is the remainder of the integral, so that
\be
A_{El}^{2} =\frac {1}{2E(I_{1}+I_{2})}.
\label{missing}
\ee
Similarly, we may write $J_{El}=A_{El}^{2}(J_{1El}+J_{2El})$.

\pr
Writing the metric functions in the form:
\bea
e^{\Gamma } & = & \left( 1-\frac {2m(r)}{r}\right) e^{-2\delta (r)}
\nn \\
e^{\Lambda }& = &  \left( 1-\frac {2m(r)}{r}\right) ^{-1},
\eea
it can easily be seen that, for spacetimes in which
\be
e^{-2\delta (r)} \equiv 1,
\ee
the integrals (\ref{jel}) and (\ref{normal}) are identical.
This is the case for a black hole without hair, i.e., the
Schwarzschild solution. In this case, $J_{El}=1/2E$,
the Hamiltonian $H$ (\ref{Hami}) is independent of $\log
\epsilon $, and the form of the partition function $Z$ is identical
to the expression (\ref{partfunct}) found in \cite{MW}.

\pr
For geometries with hair, the contribution to the integral
in (\ref{jel}) for $x\simeq 1$ is dependent on $\log
{\hat {\epsilon }}$, and is given by
\bea
J_{1El} &=  &e^{-2\delta _{h}} I_{1}.
\nn \\ & = &
\frac {A_{El}^{2}}{2E}
\left[ -\log {\hat {\epsilon }} +
 \frac {1}{2E}\sin \left(
-\frac {2Er_{h}}{1-2m_{h}'}\log {\hat {\epsilon }} \right)
\right]  .
\label{j1el}
\eea
For other values of $x$, the integrand contains a factor
of the generic form
\be
\sin^{2}
 \left( -\frac {2r_{h}E}{1-2m_{h}'} \log
{\hat {\epsilon }} \right) ,
 \cos^{2}  \left( -\frac {2r_{h}E}{1-2m_{h}'} \log
{\hat {\epsilon }} \right) ,
\ee
\be
 2\sin \left( -\frac {2r_{h}E}{1-2m_{h}'} \log
{\hat {\epsilon }} \right) \cos
 \left( -\frac {2r_{h}E}{1-2m_{h}'} \log
{\hat {\epsilon }} \right) ,
\label{factor}
\ee
which is the same for both $I_{2}$ and $J_{2}$.
Note that the logarithmic terms in the expressions
(\ref{j1el}) and (\ref{factor}) have as their argument
${\hat {\epsilon }}=\epsilon /r_{h}$ rather than $L/\epsilon $
as has been discussed by \cite{soloduk},
the contribution to the entropy for large values of $r$ being
proportional to $L^{3}$ \cite{MW}.
The reason for this is that the expansion we are using for the
integrand
in (\ref{jel}) is valid only for values of $x$ very close to 1.
An alternative expansion in inverse powers of $x$ has to be
used for large values of $x$.

\pr
We therefore write
\bea
I_{2} & = &
({}^{1}{\cal {F}}) \sin^{2}
 \left( -\frac {2r_{h}E}{1-2m_{h}'} \log
{\hat {\epsilon }} \right)
+({}^{2}{\cal {F}})\cos^{2}  \left( -\frac {2r_{h}E}{1-2m_{h}'} \log
{\hat {\epsilon }} \right) \nn \\ & &
+({}^{3}{\cal {F}}) 2 \sin \left( -\frac {2r_{h}E}{1-2m_{h}'} \log
{\hat {\epsilon }} \right) \cos
 \left( -\frac {2r_{h}E}{1-2m_{h}'} \log
{\hat {\epsilon }} \right) \nn \\
J_{2El} & = &
({}^{1}{\cal {G}}_{El})\sin^{2}
 \left( -\frac {2r_{h}E}{1-2m_{h}'} \log
{\hat {\epsilon }} \right)
+({}^{2}{\cal {G}}_{El})\cos^{2}  \left( -\frac {2r_{h}E}{1-2m_{h}'} \log
{\hat {\epsilon }} \right) \nn \\ & &
+({}^{3}{\cal {G}}_{El}) 2\sin \left( -\frac {2r_{h}E}{1-2m_{h}'} \log
{\hat {\epsilon }} \right) \cos
 \left( -\frac {2r_{h}E}{1-2m_{h}'} \log
{\hat {\epsilon }} \right)
\eea
where the ${}^{i}{\cal {F}}$'s and ${}^{j}{\cal {G}}$'s are constants
which are independent of ${\hat {\epsilon }}$ but do depend on $L$.
In total, therefore, we have
\bea
J_{El}& =& \frac {1}{2E}
\frac {J_{1El}+J_{2El}}{e^{2\delta _{h}}J_{1El}+I_{2El}}
\nn \\
& = & \frac {1}{2E} \left[ e^{-2\delta _{h}}+
\frac {J_{2El}-I_{2El}e^{-2\delta _{h}}}{e^{2\delta
_{h}}J_{1El}+I_{2El}}
\right]
\label{Jfinal}
\eea
The second term in this expression is of the order of
$(\log {\hat {\epsilon }})^{-1}$, and hence is very small.

\pr
We now proceed to evaluate the rate of change of $J_{El}$ with
respect to $\log {\hat {\epsilon }}$. The result is:
\be
\frac {\partial J_{El}}{\partial \log {\hat {\epsilon }}}
=\frac {1}{2E}
\frac {1}{e^{2\delta _{h}}J_{1El}+I_{2El}}
\left[ \frac {\partial J_{2El}}{\partial \log {\hat {\epsilon }}}
-e^{-2\delta _{h}}
\frac {\partial I_{2El}}{\partial \log {\hat {\epsilon }}}
\right] + {\mbox {smaller terms}}.
\label{jepsdep}
\ee
We therefore arrive at the following expression for the
$\epsilon $-dependence of the scalar-field Hamiltonian H:
\be
\frac {\partial H}{\partial \log {\hat {\epsilon }}}
=\frac {1}{2} \sum _{E,l,m} (a_{El}a_{El}^{\dagger }
+a_{El}^{\dagger }a_{El}) E^{2}
\frac {\partial J_{El}}{\partial \log {\hat {\epsilon }}}.
\label{Hepsdep}
\ee

\pr
To estimate the order of magnitude of (\ref{Hepsdep}), we
need to estimate the coefficients ${\cal {F}}$ and ${\cal {G}}$.
It is not possible to do this analytically, but it is
straightforward to see that
\be
I_{2}, J_{2El} \sim
{\cal {F}},{\cal {G}} \sim \frac {r_{h}}{\mu }
\label{Gepsdep}
\ee
multiplied by a numerical factor of order unity. In addition we have
\be
\frac {\partial J_{2}}{\partial \log {\hat {\epsilon }}}
\sim r_{h}\mu  J_{2}
 \sim r_{h}^{2}, \qquad
\frac {\partial I_{2}}{\partial \log {\hat {\epsilon }}}
\sim r_{h}\mu  I_{2}
 \sim r_{h}^{2}.
\ee
The leading order term in the denominator of (\ref{jepsdep})
is $J_{1}\sim \mu ^{-1}\log {\hat {\epsilon }}$
whence
\be
\frac {\partial J_{El}}{\partial \log {\hat {\epsilon }}}
\sim \frac {1}{\mu }\frac {1}{J_{1}} r_{h}^{2}
\sim r_{h}^{2} (\log {\hat {\epsilon }})^{-1}
\ee
and our final estimate is
\be
\frac {\partial H}{\partial \log {\hat {\epsilon }}}
\sim \mu ^{2} r_{h}^{2}(\log {\hat {\epsilon }})^{-1}
\label{finalres}
\ee
up to a numerical factor of order unity.
This dependence on $\log {\hat {\epsilon }}$ will be
negligible in the regime where $\epsilon \rightarrow 0$.
As we discuss in more detail in the next section, this
provides an important justification for the adiabatic
approximation used in this paper, which is the context
in which we interpret the logarithmic
divergences in the partition function (\ref{partfunct})
and the entropy (\ref{NES}).

\pr
We complete this section by commenting on
the relationship of this analysis to that of \cite{MW}.
We are fortunate in that the terms in $H$ which are dependent
on $\log {\hat {\epsilon }}$ are very small, leaving in effect
\be
H=- \frac {1}{4} \sum _{E,l,m} \left(
a_{El}a_{El}^{\dagger }+a_{El}^{\dagger }a_{El}
\right)  E \left(1+e^{-2\delta _{h}} \right)
\label{Hone}
\ee
which is to be compared with the expression
\be
H=- \frac {1}{2} \sum _{E,l,m} \left(
a_{El}a_{El}^{\dagger }+a_{El}^{\dagger }a_{El}
\right)   E
\label{Htwo}
\ee
which was employed in \cite{MW}.
The only difference between
these two expression is the $e^{-2\delta _{h}}$
term.  As discussed in \cite{MW}, the WKB approximation used
is only valid for black-hole solutions with large radii
$r_{h}$, for which $\delta _{h}$ is very small.
The two expressions (\ref{Hone}) and (\ref{Htwo}) are approximately
the same in this regime.

\section{Modified Quantum Liouville Equation}

\pr
We now turn to the interpretation of the logarithmic divergences
in the partition function (\ref{partfunct}),
in the entropy (\ref{NES}), and
in the effective Hamiltonian (\ref{finalres}). As usual, we assume
that the density matrix for the scalar field is given by the
Gibbs formula:
\be
\rho = \frac {e^{-\beta H}}{Z}
\label{Gibbs}
\ee
where $\beta$ is the inverse of the effective temperature of
the EYM black hole given in (\ref{partfunct}).
The denominator in (\ref{Gibbs}) ensures probability conservation:
$\hbox{Tr}\rho = 1$ at all times.
One would expect (\ref{Gibbs}) to be valid (at least
approximately), if the black hole is (at least approximately)
static, as we assume in the adiabatic approximation used in
this paper. However, we emphasize that we do not have a proof of
(\ref{Gibbs}) in this particular case. If it is subject to
corrections, there may be additional corrections to the
quantum Liouville equation for $\rho$, beyond the one we
exhibit shortly.

\pr
If the Gibbs formula (\ref{Gibbs}) were exact, we would normally
conclude that $\partial \rho / \partial t = 0$,
since $[\rho, H] = 0$. However, we know from the analysis of
Hawking \cite{decay} that quantum effects cause the black hole
to decay, changing its effective radius,
c.f., (\ref{NES}):
\bea
r_{q}^{2}&  =&
\left( \frac {1}{G_{N,0}}+
\frac {(1-2m_{h}')e^{-3\delta _{h}}}{360 \pi r_{h} \epsilon }
\right)G_{N,ren}   r_{h}^{2} \nn \\ & &
- \frac {G_{N,ren}}{\pi }\left[
\frac {1}{180} \left( 2-4m_{h}' +\frac {m_{h}''}{r_{h}}
\right) e^{-3\delta _{h}} -
\frac {1}{12} r_{h}^{2} \mu ^{2} e^{-\delta _{h}} \right]
\log \left( \frac {\epsilon }{r_{h}} \right)
\label{changes}
\eea
The fact that the radius changes as the black hole decays,
i.e., changes with time, motivates the identification of
time $t$ with $\hbox{log} \hat \epsilon$.

\pr
This identification
was supported in the two-dimensional string case by explicit
calculations \cite{EMN}, for example of the black-hole metric
\cite{Wbh}. In that case, we were able to incorporate the
effect of back reaction using world-sheet instantons and the
Liouville field, but here we do not have the corresponding
formal tools at our disposal. Nevertheless, amplitudes and
correlation functions in the four-dimensional case exhibit
logarithmic divergences in complete formal analogy with the
two-dimensional case, supporting the identification of
$\hbox{log} \hat \epsilon$ with time.

\pr
Since both the Hamiltonian $H$ (\ref{Hepsdep})
and the partition function $Z$ (\ref{partfunct}) depend
on $\hbox{log} \hat \epsilon$, this identification implies
that, even assuming the Gibbs formula (\ref{Gibbs}), there
are possible sources of time variation in $\rho$ which
are not given by the usual commutator term $[\rho, H]$:
\be
\frac {\partial \rho }{ \partial t} =
\left[ -\beta \frac {\partial H}{\partial t}
+\beta \frac {\partial F}{\partial t} \right] \rho
\label{rhotdep}
\ee
where the free energy $F$ is calculated in \cite{MW}.
Since the normal commutator term $[\rho, H]$ vanishes, the
result (\ref{rhotdep}) implies a modification of the quantum
Liouville term by the addition of a non-commutator term as in
(\ref{ehns}), with
\bea
&~&\nd{\delta H} = \beta \left(- \frac {\partial H}{\partial t}
+\frac {\partial F}{\partial t} \right)
=
O[(\hbox{log} \hat \epsilon)^{-1}]~+ \nn \\
&~&\frac {1}{1440 }(1-2m_{h}')e^{-3\delta _{h}}
\frac {1}{{\hat {\epsilon }}}
+\frac {1}{720}\left( 2-4m_{h}'+\frac {m_{h}''}{r_{h}} \right)
e^{-3\delta _{h}}
-\frac {1}{24} r_{h}^{2} \mu ^{2} e^{-\delta _{h}}
\label{Hslash}
\eea
which is our main result. We note that the first term,
obtained from the Hamiltonian ({\ref{Hami}), is suppressed relative
to the terms obtained from the free energy. Such a suppression is
necessary for the consistency of our adiabatic approximation.
The term in (\ref{Hslash}) that depends
on the energy content of the scalar matter field
is the last one, which is of $O[\mu ^2]$, where $\mu$ is the
mass of the scalar particle.

\pr
The rest of the terms depend on details of the
black-hole background, and in particular on its
hair~\cite{MW}. For extreme macroscopic black holes
with $m_h >> 1$ (in units of the Planck mass $M_P$), for instance,
the dominant term is the $\mu ^2$ term. This appears not to be
the case
for non-extreme black holes, where the usual
Bekenstein-Hawking term seems to be the dominant one, for $\epsilon
\rightarrow 0$. However, such terms may always be absorbed
in a renormalization of the gravitational
coupling constants of the model~\cite{soloduk,MW}, as
explained at the end of section 2. Therefore, even in this case,
it is the matter $\mu ^2$ term in (\ref{Hslash}) that
determines the order of magnitude of the modifications of
quantum mechanics~\cite{EHNS}.

\pr
In general, the quadratic dependence of $\nd{\delta H}$ on the
scalar mass $\mu$, divided in order of magnitude by just one
power of $M_P$, is the largest that could be expected for any
such modification of the quantum Liouville equation:
{\it a priori}, it could have been suppressed by one or more
additional powers of $\mu / M_P$, or an exponential, or
even absent all together. We are not in a position
to estimate the coefficient of this $\mu^2/M_P$ term.
Nor, indeed, can we be sure that such a parametric dependence
would persist in a complete quantum theory of gravity. However,
we would like to remind the reader that just such a quadratic
dependence also appeared as a possibility in a previous string
analysis \cite{EMN} of the quantum Liouville equation. We
cannot resist pointing out also that such a dependence may not be
many orders of magnitude from the experimental sensitivity to
such a modification of the quantum Liouville equation for the
$K^0$ system \cite{EHNS,EMN,CPLEAR}.

\pr
Notice also that the main time-dependence in (\ref{Hslash})
comes from the divergences occurring when one traces over
an infinite number of
states in $Z$. Any explicit time-dependence of the
Hamiltonian operator $H$ is subleading at large times (\ref{finalres}).
This is not in contradiction with the adiabatic approximation
made above, in which the effect of quantum fluctuations
on the Hawking temperature $\beta$ was neglected.
This approximation is valid for macroscopic black holes,
for which the $WKB$ approximation method we have used to derive
expressions for the entropy and the free energy
is valid~\cite{MW}.

\pr
It is easy to check that the
$\epsilon$-dependence in (\ref{Hepsdep}), combined with the
Gibbs formula (\ref{Gibbs}), reproduces the $\epsilon$-dependence
(\ref{NES}) of the entropy found previously in \cite{MW}.
The interpretation developed earlier
of the $\epsilon$-dependence as a time-dependence implies that
the entropy of the scalar field $\phi$ increases with time:
\be
\frac {\partial S}{\partial t} =
-\frac {1}{360} (1-2m_{h}') e^{-3\delta _{h}}
\frac {1}{{\hat {\epsilon }}}
-\frac {1}{180} \left( 2-4m_{h}' +\frac {m_{h}''}{r_{h}} \right)
e^{-3\delta _{h}}
+\frac {1}{12}r_{h}^{2} \mu ^{2} e^{-\delta _{h}}
\label{Sdot}
\ee
Thus the $\phi$ state evolves to become more mixed, as a result of
the non-commutator term (\ref{Hslash}) in the quantum
Liouville equation (\ref{ehns}). This can be thought of as being
due to a change in the entanglement of the external $\phi$ field
with the unmeasurable modes interior to the black hole.
The amount of this entanglement depends on the quantum numbers
of the EYM background, as can be seen in (\ref{NES}). The more
hair it has, the smaller the $\epsilon$-dependence in (\ref{Hepsdep}),
and, correspondingly, the slower the rate of information
loss in (\ref{Sdot}).
\pr
Although our analysis breaks down in the
limit of an extremal black hole, the results
(\ref{Hslash}) and (\ref{Sdot}) suggest that there may be
information loss even in this case.
This is because it is no longer true in general at the
quantum level that the entropy is proportional to the
area of the horizon (\ref{entr}). There will be
information loss (entropy increase) even in the absence
of any finite-temperature effects, if there
is entanglement with modes beyond the horizon at the
quantum level, as we have illustrated.
This observation is related to the phenomenon of entropy
generation during inflation in cosmology, which may also
be regarded as a non-equilibrium process associated
with information loss beyond the Hubble horizon.
In the context of non-critical strings~\cite{EMN}, we have
discussed in ref. \cite{emninfl} how such an information loss
can lead to a stochastic framework for time evolution,
during such non-equilibrium processes.
The stochasticity of the time evolution
in the Liouville string~\cite{emninfl}, where the
time variable is identified with a RG scale,
can be derived from some specific
properties of the RG evolution in two-dimensional spaces
(world sheets)~\cite{friedan,EMN}.
One can hope that a similar framework may be developed here,
identifying a renormalization group (UV) scale
$\hbox{log} \hat \epsilon$ with time.
However, we are not yet in a position to prove that
a similar stochasticity characterizes the
RG evolution in the four-dimensional case.
However,
the presence of logarithmic
infinities in the (entanglement) entropy of black holes does
seem to be a generic phenomenon,
independent of the dimensionality of space-time, in view of
the fact that they are present even in
two-dimensionsal models~\cite{uglum}.

\pr
The above analysis is not complete yet,
as a result of the absence
of a satisfactory treatment of quantum gravity.
In the absence of such a treatment, the results of this paper can
only be regarded as indicative. However, we think that they
constitute interesting circumstantial evidence in favour of the
picture advanced previously \cite{Hawking, EHNS, EMN}, namely
that microscopic quantum fluctuations in the space-time background
may induce a loss of quantum coherence in apparently isolated
systems. Moreover, the magnitude of $\nd{\delta H}$ that we find
is consistent with previous string estimates, and may not lie
many orders of magnitude beyond the reach of particle physics
experiments \cite{CPLEAR}.
\pr
\nk {\bf Acknowledgements}
\pr
The work of D.V.N. is supported
in part by D.O.E. Grant
DEFG05-91-GR-40633.
E.W. thanks EPSRC (U.K.) for a research studentship.

\end{document}